*Review*

# An Overview on the Landscape of Self-Adaptive Cloud Design and Operation Patterns: Goals, Strategies, Tooling, Evaluation, and Dataset Perspectives

**Apostolos Angelis and George Kousiouris ***

Deptartment of Informatics and Telematics, Harokopio University, 17778 Athens, Greece; aangelis@hua.gr
* Correspondence: gkousiou@hua.gr

**Abstract**

Cloud-native applications have significantly advanced the development and scalability of online services through the use of microservices and modular architectures. However, achieving adaptability, resilience, and efficient performance management within cloud environments remains a key challenge. This work systematically reviews 111 publications from the last eight years on self-adaptive cloud design and operations patterns, classifying them by objectives, control scope, decision-making approach, automation level, and validation methods. Our analysis reveals that performance optimization dominates research goals, followed by cost reduction and security enhancement, with availability and reliability underexplored. Reactive feedback loops prevail, while proactive approaches—often leveraging machine learning—are increasingly applied to predictive resource provisioning and application management. Resource-oriented adaptation strategies are common, but direct application-level reconfiguration remains scarce, representing a promising research gap. We further catalog tools, platforms, and more than 30 publicly accessible datasets used in validation, and that dataset usage is fragmented without a de facto standard. Finally, we map the research findings on a generic application and system-level design for self-adaptive applications, including a proposal for a federated learning approach for SaaS application Agents. This blueprint aims to guide future work toward more intelligent, context-aware cloud automation.

## 1. Introduction

Cloud computing introduced new architectural approaches to application development; monolithic applications evolved to multiple smaller loosely coupled components or services [1]. This collection of independent services that communicate through lightweight APIs offers several benefits, including agility, adaptability, scalability, and performance improvements [2,3]. Following many big tech firms, the industry migrated to this new approach.

This trend introduced several challenges, as cloud-native applications depend upon complex distributed architectures as well as dynamic and multitenant infrastructure layers. This dynamicity and distributed nature in many cases increases the risk of failure and dictates the need for constant monitoring and adaptation mechanisms. Cloud-native applications also brought several improvements to online service development, including high maintainability and scalability [4].

In order to aid in their creation, design, and operation patterns, best practices have been documented by practitioners, setting the industry standard in how to mix and use multiple small and independent services [5,6]. To define the concept of a pattern, one can follow the generic definition included in [7], which mentions that a pattern is “a proven series of activities which are supposed to overcome a recurring problem in a certain context, particular objective, and specific initial condition”. One key aspect of patterns is the fact that choosing their parameters can significantly determine whether the pattern is beneficial or harmful. Thus, specific caution and automation must be applied in order to ensure that the former is achieved.

Patterns or strategies may be applied either at the application design or architecture level, regarding its structure, or even at the management level, e.g., targeting at making decisions during runtime. The latter may involve decisions on the amount of resources assigned to an application, the location of its execution, the type of resources needed, etc., thus being characterized as cloud operations patterns aiming to enhance performance and cost. Others, such as the compute resource consolidation pattern [8], are dictated by the inherent multitenancy of the cloud computing model.

The aim of this work is to systematically record and analyze the literature published in the last eight years concerning the aforementioned issues of cloud design and operation patterns and strategies. More specifically, it studies the context (e.g., application, network, infrastructure), the intended problem (e.g., resource assignment, execution location, etc.), the series of activities (types of mechanisms for automation), the initial condition (needed data and tools), as well as the overall objective (cost, performance, reliability, etc.). These are formulated as five research questions (detailed in Section 3) as well as a taxonomy of characteristics upon which related works are mapped to. Through this, we highlight the current status of the field and identify potential future directions.

From a practical point of view, a number of issues are examined, in terms of the usage of software (such as programming languages, frameworks, tools), platforms, and datasets in the examined pattern mechanisms. Especially for datasets, an analysis of their contents is included, in order to aid future researchers in an easier selection of the appropriate source. Additionally, the experimental processes and the degree and method of validating the experimental results are examined.

Furthermore, based on the overview findings, we propose an application and system-level design for cloud self-adaptive applications. The components of this blueprint are specifically annotated to demonstrate their connection to the research questions we have investigated. This proposal embeds intelligent adaptation mechanisms directly within the application and system structure, as well as collaboration between discrete application instances.

The remainder of this paper is structured as follows: Section 2 introduces the background. Section 3 introduces the motivation, the research questions, and compares this overview with other works. Section 4 presents the research methodology and Section 5 includes the categorization and findings for the overviewed works. Finally, Section 6 provides a discussion on the research questions, while Section 7 concludes the paper and sketches future research directions.

## 2. Background

Cloud Design Patterns and Cloud Operation Patterns are architectural and operational solutions for reoccurring problems encountered when building and managing applications in cloud environments. Cloud Design Patterns address architectural concerns such as data distribution, service orchestration, and fault tolerance, while Cloud Operation

Patterns focus on optimizing system operations through automation, monitoring, and incident response.

### 2.1. Cloud Design Patterns

The majority of cloud software and services relies on a modular microservice architecture, a proven choice for improved developer productivity and optimal selection of the most appropriate combination of technologies, but with the expense of increased complexity. This level of complexity of the cloud software and services architecture makes reusability and automation a prerequisite [9].

To facilitate the development of cloud-native applications, seasoned practitioners have devised architectural patterns that encapsulate their expertise in resolving recurring issues [8]. Software and services' design patterns are like blueprints for common design problems; they provide reusable solutions that can be adapted to different situations. Using these patterns, developers can create more efficient, reliable, and maintainable solutions, learning from past successes and avoiding common pitfalls.

A range of microservice-level architectural patterns have been investigated and proposed; these include Log Aggregator (aggregation of distributed logs in a central location and subsequent root cause analysis) , Batch Request Aggregator [10] (based on model-driven consolidation of incoming requests to ease back-end stress), Service Registry, Service Discovery, Circuit Breaker (with intelligent and dynamic transition between the circuit states), API Gateway, and Health Check, among others [11]. One of the interesting aspects to be investigated in this case is how these mechanisms can be combined with intelligent approaches for dynamically adapting to the current conditions an application has to face, avoiding static configurations.

### 2.2. Cloud Operation Patterns

Cloud infrastructures, due to their large scale, complexity, geographical distribution, and heterogeneity, require extensive configuration and fine-tuning by administrators; these laborious tasks have been investigated in order to be automated. The initial approach to tackle the complexity of administrative tasks was to employ methods such as benchmarking, statistical models, time-series analysis, threshold-based policies, and heuristic algorithms [12]. Subsequently, artificial neural networks (ANNs) [13] optimized by genetic algorithms (GAs) [14,15] were studied and applied to design rationale and predictive management procedures. Prior research explored multiple methods managers can utilize to reap the rewards of machine learning (ML) [16].

ML has shown potential to make proactive decisions in multiple parts of cloud computing infrastructures, such as energy consumption optimizations [17]. Container scheduling, server defragmenter/migration manager, power capping manager, and server health manager are just a fragment of the extensive opportunities for ML-driven resource management [12]. Furthermore, strategies that enhance the resilience of cloud systems against failures have been highlighted in recent studies [18].

Key infrastructure-level self-adaptive patterns are self-healing (automatically detect and repair faults or failures in infrastructure components), auto-scaling (dynamically adjust the number of resources based on demand), resource optimization (efficiently allocate and utilize resources to minimize costs and maximize performance), placement optimization (selection of complementary applications as node neighbors), cloud-edge collaboration and workload distribution, predictive analysis (use historical data and machine learning to forecast future resource needs and proactively adjust capacity), security automation (automatically detect and respond to security threats, such as intrusion attempts or mal-ware), and monitoring and detection (proactively observe and identify issues within cloud environment) mechanisms.

## 3. Related Reviews, Motivation, and Research Questions

Several reviews have examined self-adaptive systems. However, these works are often narrow in scope and scarcely address objectives, control scopes, decision-making, tools, and validation in a unified way. To fill these gaps, this work formulates five research questions that structure our analysis across adaptation objectives, control mechanisms, decision-making approaches, supporting tools, and validation practices, providing both conceptual and practical perspective.

### 3.1. Related Work and Broader Perspective Motivation

A number of recent overviews and surveys exist that deal with self-adaptive approaches in specific fields of software and service development. Some of them focus on self-adaptive systems from the perspective of the technology used, such as ML-based automated systems, while others show interest in a specific field of application, such as IoT or fog computing. More specifically, Gheibi et al. [19] document more than a hundred studies with the emphasis on ML-powered automated systems with MAPE-K feedback loops that are supported by a machine learning mechanism. The study classifies the main problems that ML tries to solve and identifies the most common methods used for tasks like prediction, classification, and reinforced learning. Challenges and limitations when using ML in self-adaptive systems are also noted.

Cardellini et al. [20] focus on the algorithms used to control the adaptation of container-based applications on fog and cloud computing and more specifically on self-adaptation with respect to workload changes. Alfonso et al. [21] analyze the adaptation strategies in response to dynamic events that impact the QoS of IoT systems. This work reviewed 39 studies that mainly focus on optimizing resource consumption, QoS violation avoidance, and software update deployment patterns of IoT infrastructures.

Krupitzer et al. [22] identified 24 relevant papers on design patterns for self-adaptive systems. This work outlined seven categories (monitor, analyze, plan, execute, component structure and interaction, knowledge management, and coordination) and 55 design patterns that can be applied in IoT environments.

Kirti et al. [23] categorize various fault-tolerant techniques into four categories, reactive, proactive, adaptive, and hybrid, and analyze the fault-tolerance approaches. The survey also discusses the trade-off between lightweight predefined and heavy proactive self-adaptive techniques. Quin et al. [24] perform a study in research on decentralization of self-adaptation. The work analyzes the components and coordination mechanisms of decentralized self-adaptive systems, and identifies three coordination patterns used in the cases studied. The study concludes with the challenges for future work on decentralized self-adaptive systems. Chen and Bahsoon [25] provide an extensive taxonomy for cloud autoscaling systems. This work offers in-depth analysis of intelligent autoscaling functionality in cloud environments and outlines future research directions in this field. The taxonomy provides a foundation for building more intelligent autoscaling systems.

As shown in Table 1, prior reviews have largely concentrated on specific technologies (e.g., ML-based MAPE-K loops), specific infrastructure scopes (e.g., containers, autoscaling, IoT), or specialized objectives (e.g., fault tolerance, decentralization). In contrast, the present work offers a comprehensive cross-domain perspective, simultaneously covering both cloud design and operation patterns across application- and infrastructure-level contexts, and introduces a unified taxonomy that spans objectives, control scope, decision-making, automation level, validation methodology, and practical tooling. Furthermore, this work uniquely catalogs 38 publicly accessible datasets, along with platforms, tools, and validation strategies, enabling reproducibility and comparative experimentation.

We have also included a column 'Pending Questions' in this table, which effectively indicates open questions that have not been covered in the related surveys and were used as an inspiration for our research questions.

**Table 1.** Comparison of related reviews and core issues from the current work.

| Authors | Focus Area | Key Points/Contributions | Core Contribution | Pending Questions |
| --- | --- | --- | --- | --- |
| Gheibi et al. [19] | ML-powered automated self-adaptive systems (MAPE-K feedback loops) | Documents 100+ studies using ML in self-adaptive systems; classifies ML tasks (prediction, classification, reinforcement learning); notes challenges and limitations. | Focuses only on ML-based mechanisms and MAPE-K loops. | What are other techniques that may be used and are non-ML based? What are the datasets used for the validation process of the various algorithms? |
| Cardellini et al. [20] | Adaptation algorithms for container-based applications (fog and cloud) | Documents 42 studies; focuses on self-adaptation to workload changes in container deployments for fog/cloud. | A brief reference is made to the simulation software. | What method was used to create the load in the simulation environment? Were any additional tools or datasets involved? |
| Alfonso et al. [21] | IoT systems adaptation strategies | Reviews 39 studies optimizing resource use, avoiding QoS violations, and software updates in IoT. | Focused on IoT-specific adaptations. | What are the decision-making mechanisms and validation practices? |
| Krupitzer et al. [22] | Design patterns for self-adaptive systems (IoT) | Identifies 24 papers, 55 patterns; classifies into 7 categories (monitor, analyze, plan, execute, component structure and interaction, knowledge management, coordination). | Concentrates on IoT environments | Which are the automation levels, tooling, or dataset-driven validation methods? |
| Kirti et al. [23] | Fault-tolerance techniques | Categorizes 146 papers into reactive, proactive, adaptive, and hybrid; examines trade-offs between techniques. | Focuses specifically on fault tolerance. | What are the performance, cost, and security compromises involved? Which load generation approaches were employed, and what is the structure of the datasets? |
| Quin et al. [24] | Decentralization in self-adaptation | Analyzes decentralized components and coordination in 14 studies; identifies 3 coordination patterns; future challenges. | Decentralization-specific | Which are the performance, cost and security metrics? What tools, experimental setups and datasets would be most suitable for implementing and testing self adaptive patterns? |
| Chen and Bahsoon [25] | Cloud autoscaling system taxonomy | Provides detailed taxonomy for intelligent autoscaling; outlines future research; foundation for smart autoscaling systems. | Autoscaling-specific with a short overview of infrastructure, tools and dataset evaluation. | Can the findings be generalized beyond autoscaling to encompass multiple objectives? What tools, experimental setups, and datasets would be most suitable for implementing and testing Autoscaling Systems? |

### *3.2. Research Strategy*

We have attempted to follow the categorization levels of Chen and Bahsoon [25], as we believe that it aligns closely with the focus of our current work.

The top-level fields of the taxonomy are defined in Figure 1. The pattern's goal field captures the objectives of adaptation, such as performance optimization, cost reduction, or security enhancement, representing the reason behind adaptation. The adaptation strategy field identifies the core adaptation mechanism, from plain detection, resource management to application-level reconfiguration. The decision-making approach reflects the techniques used for making adaptation decisions. The automation-level field addresses the level of automation, ranging from manually trained systems to fully autonomous continuous adaptation, which indicates the maturity of solutions. The validation method field classifies by how research is validated (simulation, emulation, real-world experiments) and what tools/datasets are used (CloudSim, Kubernetes, Azure, Google traces, etc.). Finally, the feedback loop field defines the operational backbone of self-adaptive cloud systems, ensuring that objectives, scopes, and decision-making mechanisms are continuously connected. The lower-level ones are then populated in Section 5 from the grouping of the respective approaches identified in the overviewed works.

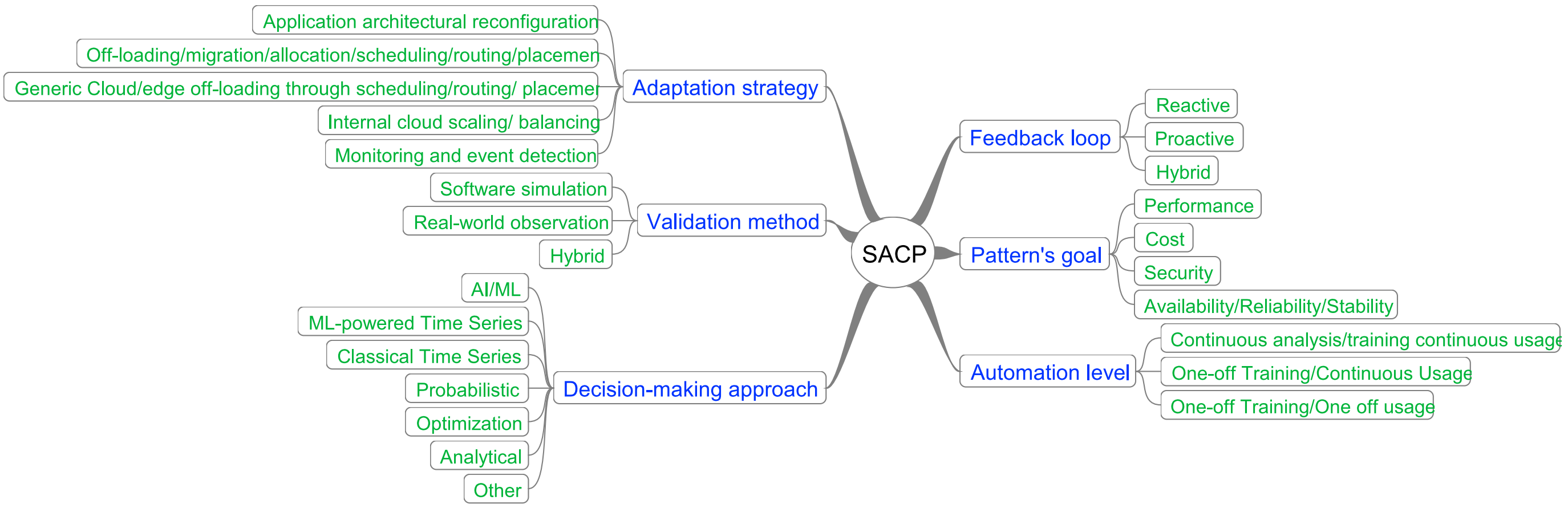

**Figure 1.** A taxonomy of Self-Adaptive Cloud Patterns (SACP) research.

Thus, the overall research questions of this work can be defined as follows:

- RQ1: Which are the patterns' objectives in cloud automation?
- RQ2: What are the scopes of control (i.e., target of regulation) of self-adaptive patterns?
- RQ3: What are the approaches used for decision-making?
- RQ4: Which software and tools were used to create the management mechanisms in the overviewed work?
- RQ5: Which methods, datasets, and tools were used for experiment validation?

## 4. Research Methodology and Defined Taxonomy

The research methodology included an initial search round to identify related work. Then, a top-level taxonomy was created (Section 4.2) to map concepts to the generic pattern definition elements (context, problem, activities, etc.). The following presents details of each step.

### 4.1. Search Strategy

Initially, the research papers were retrieved through Google Scholar. The search term was "cloud computing" AND "self-adaptive" AND "pattern", with the publication years limited to 2018–2025. Boolean operators and the year filter were intentionally chosen to be broad enough to encompass the relevant domain.

Google Scholar returned more than 11,800 papers; the top 800 entries in the results were retained for further processing. The inclusion criteria were papers published in journals or conferences, papers written in English, and papers discussing self-adaptive mechanisms related to cloud computing. Non-peer-reviewed articles were not excluded, on the grounds that the primary factor for inclusion was the paper's actual content, specifically its relevance to cloud self-adaptive patterns, even if the title or abstract did not perfectly match the defined keywords. Our selection process involved an initial filtering phase in which we reviewed the title, abstract, keywords, and conclusions of each paper. Following this, we carefully read the full text of the included papers to determine their precise relevance to our defined scope. We also examined the reference lists of included papers to identify any relevant studies potentially missed in the initial search.

A total of 111 articles were reviewed, published between 2017 and the first half of 2025; Figure 2 shows the number of articles by published year and the percentage of each type of document; we observe that the vast majority of articles (73.9%) belong to journal publications, with conferences followed with 21.7% percent. In recent years, a clear trend has been observed, indicating a very significant increase in the investigation of related topics.

In addition, to detect the status in previous years to this interval, a relevant search was performed for the period 2010 to 2017. In total, 34 according publications were discovered, indicating that there was activity on the topic in the period, although not in the volume and intensity observed over the recent years, especially from 2023 and on.

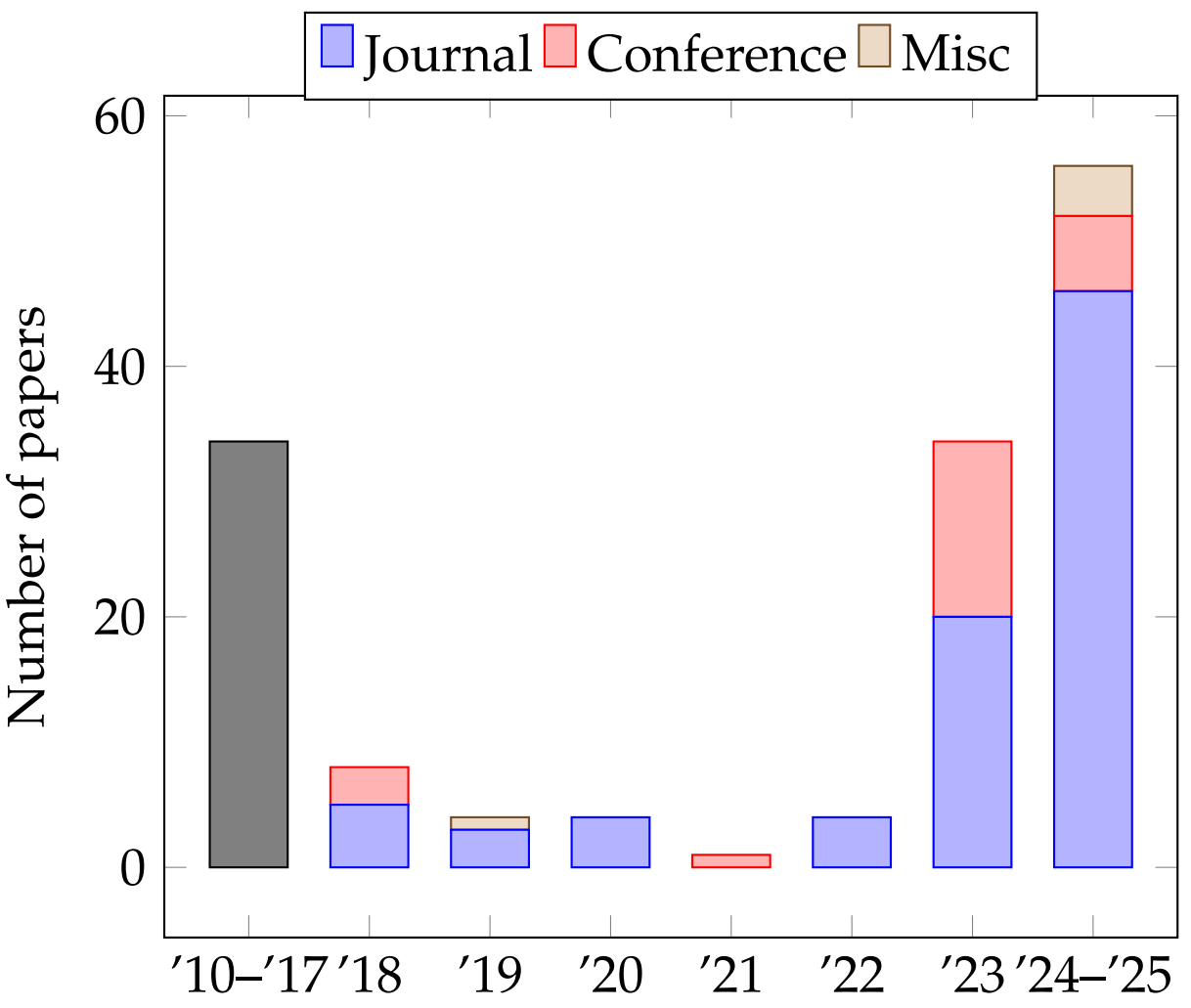

**Figure 2.** Distribution of academic papers by publication year (2010 to 2025).

### *4.2. Top-Level Taxonomy Fields and Mapping to Pattern Definition*

In order to drive the selection of the top-level taxonomy fields, one can start from the definition of a pattern, mentioned in Section 1. Starting from the objective, one needs to determine what is the primary goal of each pattern, i.e., what it tries to optimize from a non-functional perspective. Then, the series of activities can refer to the relevant decision-making approach used in the context of the pattern. The specific mechanism is typically used to solve a specific recurring problem based on an adaptation strategy in order to control a given entity. How the mechanism is applied can be also considered as part of the initial conditions as well as the recurrence of the problematic situation, indicating its automation level as well as the necessary feedback loops applied. Finally, the mechanism needs to prove its usefulness, thus needing a validation method concept.

## 5. Categorization of Related Work and Definition of Lower Levels of the Taxonomy

In this section, the investigated works are classified in subcategories of each top-level field based on our review process. It needs to be stressed that all 111 works are included in the first six tables of the top-level goals. The follow-up tables of implementation details (e.g., used software, datasets, etc.) may not include all of them, based on the information available in each work. The lower levels of the taxonomy also appear in Figure 1.

### 5.1. Pattern's Prime Goal

Most studies tackle the issue of performance optimization (Table 2), while cost reduction is the second most popular category. Performance targets include system optimization, network performance, response time reduction, network bottleneck detection, SLO/QoS violation mitigation, QoS prediction, and more.

The cost category includes energy consumption, deployment, and operating costs, while security refers to security-centered resource provisioning, fault detection, cybersecurity risk assessment, data privacy, intrusion, DDOS, and cyberattack anomaly detection systems.

A few papers target availability, reliability, and stability, typically referring to cloud monitoring combined with system anomaly detection, architectural stability, microservice circuit breaking, big service management, resilience enhancement and data privacy, workflow management, and network traffic forecasting.

**Table 2.** Classification of pattern's goal.

| Pattern's Goal | Used in Papers |
|---|---|
| Performance | [26–104] |
| Cost | [28,33,40,45,46,55–57,64,65,69,71–73,75,76,87,88,90,91,105–113] |
| Security | [46,114–130] |
| Availability/ Reliability/Stability | [85,95,110,121,131–136] |

### 5.2. Feedback Loops

A feedback loop is a cyclical process in which the output of a system is fed back as input, influencing the future behavior of the system. It is an essential part of a self-adaptive process and can be distinguished in two main categories: reactive and proactive. In the reactive case, the system collects real-time data on performance, security incidents, power usage, etc., to perform adaptation decisions that may or may not be based on a trained model, once an event that requires intervention is identified; in the proactive case, the approaches use real-time data to predict future trends and anticipate changes based on a historically trained model or Agent.

About half (59 cases) of the articles use the reactive approach to deal with state changes in the cloud application or system, an intuitive solution to system automation. Proactive approaches (51 cases) are in the general case computationally heavy and usually rely on machine learning technologies and time series analysis. The classified articles are listed according to this categorization in Table 3. One of the reviewed works [53] proposes both approaches; therefore, it is classified as hybrid.

**Table 3.** Classification of feedback loop pattern.

| Feedback Loop | Used in Papers |
|---|---|
| Reactive | [26,28,30,33,35,37,45,47,50,52,54–57,59–61,63–69,71–73,75,79–82,84,85,87,88,90–93,95,96,105–109,111,112,114–116,118,119,122,124,131–133] |
| Proactive | [27,29,31,32,34,36,38–44,46,48,49,51,58,62,70,74,76–78,83,86,89,94,97–104,110,113,117,120,121,123,125–130,134–136] |
| Hybrid | [53] |

### 5.3. Adaptation Strategy

Several different solutions have been applied for the adaptation strategy (Table 4). The majority of these resort to techniques such as migration, resource allocation, scheduling, scaling, and balancing of resources, in either one or more locations. All of the above target primarily the resource size used by an application or the locality of these resources, based on examined workload, execution time, or resource usage. On the contrary, only a few studies consider the reconfiguration of applications, i.e., actions that alter in a way the inner behavior or architecture of an application in order to better adapt to changing environmental conditions (e.g., batching requests together to be executed by a single thread in order to reduce excessive thread creation).

There are some interesting cases that appear to be using application-level context in order to drive the self-adaptive mechanism strategy. For example, in the e-health system of Karan Bajaj and Singh [72], the according scheduling or offloading of the request is based on an estimate of the criticality of a patient. Tasks from more critical patients are executed on the edge in order to reduce latency. Thus, the control mechanism digs into the application layer in order to aid in a more fine-grained performance management of the requests between the edge and the cloud. In a similar case, Tundo et al. [64] present an energy-aware approach for self-adaptive AI-based applications that can balance accuracy in object detection with energy consumption. In this case, the image analysis uses either CPU- or GPU-based resources based on each image's complexity characteristics and needed accuracy of detection. One last case is the work in Yin et al. [54], which uses real-world mobile device trajectories in the form of time-stamped GPS information to feed mobility-aware off-and-downloading task algorithms in mobile edge computing environments. Thus, the resources used at the edge servers by a mobile application follow as closely as possible the respective user movement by offloading the needed computation tasks to the nearest edge server.

Moreover, there is a wide range of works that focus extensively on monitoring and event detection/prediction of operations, without dealing with the corrective action part of the process. They are, however, included in this classification since this part of the process is critical to a successful adaptation strategy.

**Table 4.** Classification of adaptation strategy.

| Adaptation Strategy | Used in Papers |
| --- | --- |
| Application architectural reconfiguration | [37,44,91,132,133] |
| Off-loading/placement based on application context | [54,64,72] |
| Generic Cloud/edge off-loading through scheduling/routing/placement | [26,49–52,55–57,60,61,65,67–69,71,73,75,77,80,84,86–88,90,92–95,98–100,102,104,108,109,111–113,124,135] |
| Internal cloud scaling/balancing | [27,28,30–36,39–42,45–47,53,59,61,63,66,76,78,79,81–83,85,95–97,105–107,110] |
| Monitoring and event detection (without adaptation action planning) | [29,38,39,43,48,58,62,70,74,89,101,103,114–123,125–131,134,136] |

### 5.4. Experiment Validation Methodology

Experiments require rigorous validation to ensure the reliability and reproducibility of the results. In the researched work, two approaches were employed: software simulation and real-world computing infrastructure. Each method offers distinct advantages and challenges.

Software simulation allows for precise control over experimental variables, minimizing external factors that could influence the results. It is cost-effective, enables rapid iteration, and can be easily scaled to handle varying system sizes. However, simulations often involve simplifying assumptions and abstractions of real-world systems, which may limit the accuracy and generalization of findings. Additionally, software simulations may not fully capture the physical limitations and constraints of real-world infrastructure, so validating the accuracy of simulations requires benchmarking against real-world experiments. However, this is alleviated by the fact that in most cases, the simulated experiments utilized real-world datasets, as depicted in Section 5.8.

Real-world computing infrastructure provides insights into the actual performance and behavior of systems under realistic conditions. The results are more likely to be generalizable to real-world deployments. However, setting up and maintaining real-world infrastructure can be expensive and complex. Moreover, experiments are susceptible to external factors like neighboring cloud workloads fluctuation, which can introduce variability and noise to the data. In general, real-world computing experiments are often difficult to implement in relation to simulation, a fact that is recorded in the percentage they took part among the studied literature, as shown in Table 5.

Validating experiments using a hybrid approach that combines simulation and real-world infrastructure is the exception among the researched papers.

**Table 5.** Classification of experiment validation method.

| Validation Method | Used in Papers |
| --- | --- |
| Software simulation | [28–33,37–39,39–41,43,45,46,48,50–54,56–58,60–63,65,67–75,77–80,82–84,86,88,90,93–98,100–109,111,112,114–116,118–130,133–136] |
| Real-world observation | [26,27,34–36,42,44,47,49,55,59,64,66,73,76,81,85,87,89,91,92,110,113,117,131,132] |
| Hybrid | [73,99] |

*5.5. Runtime Automation Level*

In relation to how the various mechanisms utilize the available data (that is, for training) and the timing in which they are consulted by the overall system, three different categories can be identified (Table 6).

By a wide margin, mechanisms that repeat the training/analysis phase continuously and also use the predicted outcomes during runtime are the most popular category. Examples of such mechanisms include typically methods such as reinforcement learning, heuristic optimization based on changing runtime conditions, etc.

Although not many, there are papers that propose mechanisms that can be used after an initial training session. These mechanisms have a one-off training process and then are used continuously during runtime to have a more adaptive system. These cases may be occasionally retrained; however, this retraining is not part of the runtime loop/process.

Lastly, there are a few cases in which training is applied in a one-off manner and decision-making is also performed in a one-off manner (i.e., during deployment time for the selection of provider or resources).

**Table 6.** Classification of runtime automation level.

| Automation Level | Used in Papers |
| --- | --- |
| Continuous analysis/training continuous usage | [26–30,32–37,39–42,44–47,49–51,53–55,57,59–71,73–87,90–93,95,97–100,102,104–108,110–112,118–121,124,125,127,130–133,136] |
| One-off Training/Continuous Usage | [31,38,39,43,48,52,56,58,72,88,89,96,101,103,113–117,122,123,126,128,129,134,135] |
| One-off Training/One-off usage | [94,109] |

### 5.6. Pattern's Decision-Making Approach

Self-adaptive software dynamically adjusts its behavior and resource allocation in response to changing conditions, and at the heart of this capability lies a decision-making process. The decision-making approach encompasses a spectrum of techniques ranging from simple rule-based triggers to complex, AI-driven predictive models.

An overview of the decision-making approaches appears in Table 7. The specific low-level categorization was kept at a coarse-grained level, not delving into subcategories of each field. This was done primarily due to the fact that there are already exhaustive overviews and surveys [19–25] that deal with this issue, as mentioned in Section 3.

The prominent decision-making method used by the publications is that of machine learning. Machine learning offers the feature of the required intelligence to deal with new, unknown situations that are going to adversely affect the application or infrastructure. Function approximation approaches may be used to understand the needed size of resources (or any other corrective action applied), typically coupled in many cases with reinforcement learning for continuous improvement. Classification cases commonly apply to security-oriented approaches for detection of illegal traffic. Another usage scenario is for resource scheduling and allocation, based on the integration of graph neural networks and swarm optimization.

A typical decision-making approach also relates to time-series analysis due to the user-centric cloud workload cyclic patterns. Time-series algorithms enable self-adaptive systems to learn from historical data and predict future trends, allowing them to proactively adjust their behavior to changing conditions. A common use case is a self-adaptive load-balancing (between locations) or auto-scaling strategy (within the same location), leveraging the temporal periodic patterns in user access to cloud services to improve performance. For this category, we have created two entries in Table 7, one relating to ML-driven approaches (e.g., LSTM architectures) and one relating to more classical time-series methods (e.g., ARIMA).

**Table 7.** Classification of pattern's decision-making approach.

| Decision-Making Approach | Used in Papers |
| --- | --- |
| AI/ML | [29,38,39,39,40,56,58,62,68,70,72,76,77,82,82,84,88, 90,92,93,96,98,101–104,110–122,125–130,133,135,136] |
| ML-powered Time Series | [46,48,53,74,78,89,123,134] |
| Classical Time Series | [27,31,32,41–43,49,95] |
| Probabilistic | [28,33,35,37,44,51,63,69,83,88,94,95,105,107,108,125] |
| Optimization | [26,30,30,35,37,39,45,47,52,54,55,57,60,61,64–67,73,75,79,80,99,100,109,124,131] |
| Analytical | [34,36,50,59,71,85,87,91,106,108] |
| Other | [81,82,86,97,109,132] |

In relation to probabilistic-based implementations, a typical one is a self-adaptive architecture to detect and manage underutilized or overloaded virtual compute resources in response to workload changes while focusing on additional criteria like the performance of the consolidation procedure. Likewise, resource allocation algorithms, using probability formulas, may target at reducing power consumption and number of migrations.

Optimization algorithms are also widely used to provide a systematic approach to finding near-optimal solutions to complex problems, especially when exact solutions are computationally intractable. Common use cases refer to identifying optimized placement schemes (e.g., deployment plans for service to physical node mapping, selection of clusters, etc.). Approaches such as genetic algorithms and swarm colony optimization may be used to dynamically fine-tune system behavior and optimize one or more features (e.g., cost and performance), while taking other parameters as constraints (e.g., resource utilization, user affinity requirements, legal requirements, network transmission overhead reduction, power consumption, etc.).

Analytical algorithms include the definition of detailed mathematical equations that describe the underlying system, which are then solved using computational algorithms. They are particularly useful for tackling self-adaptive problems that target precision at the expense of efficiency. This is due to the fact that they need a large period of time to analyze the system as well as deep knowledge of the latter to describe it accurately.

There is also a more general category that includes a variety of other approaches, including fuzzy-logic [82,109], physics-inspired system modeling [81], control-based methods [132], PID controller algorithms [86], and Algorithmic Game Theory [97].

### *5.7. Software and Tools*

Given that the practical approaches of an experimentation and validation process are in many cases the stage that is the most time-consuming, an effort was made to concentrate information that may prove to be helpful for researchers in future endeavors. To this end, we highlight the main elements of such a process, including the software and infrastructure used as well as other related tools in the investigated works.

#### 5.7.1. Infrastructure Platforms and Software

In this section, we highlight the specific platforms and software (Table 8) used in the experimentation process of the investigated works that were involved in real-world experiments, as indicated in Table 5. About a quarter of the publications (26) use computing infrastructure for the needs of experiments' implementation, either in a public or private cloud or a combination of both. Fourteen of these infrastructures refer to public cloud environments (e.g., AWS, Alibaba etc.). Seventeen cases used some form of open source platform software, in many cases combined with public cloud resources such as VMs.

This hybrid combination gives some critical benefits. More specifically, it includes inherently the variability of the multitenant public cloud infrastructure, while on the other hand, it gives the ability to intervene more in the way decision making or configuration is performed at the platform level (e.g., to investigate placement or routing in Kubernetes clusters). The most prominent of these tools is Kubernetes, used as the main container orchestration system for the experiments.

**Table 8.** Infrastructure platforms and software used in experiments.

| Name | Type | Freely Available | Used in Papers |
|---|---|---|---|
| Alibaba | public cloud | No | [42,110] |
| Amazon Web Services | public cloud | No | [26,35,42,47,66,81,87] |
| Azure | public cloud | No | [62,87] |
| Docker/Docker Swarm | OS-level virtualization with built-in orchestration system | Yes | [27,81] |
| Google Cloud | public cloud | No | [66,76,89] |
| Kubernetes | container orchestration system | Yes | [34,36,42,47,55,59,66,85,92,110,117] |
| NodeRED | flow-based visual programming tool | Yes | [87,91] |
| OpenvSwitch | software network switch | Yes | [93] |
| OpenWhisk | serverless functions platform | Yes | [87,91,92] |

5.7.2. Simulation Software Used

In this section, we highlight the software used for experiment simulation (Table 9). CloudSim was the software of choice for several researchers, while there is also a multitude of software that was used individually by various papers. There are many cases that based the experiments on custom solutions and are not included in the table. In addition, MATLAB is used in some cases as a simulation engine, but in most of them as a model creation environment. For this reason, it is included in the following subsection. In all cases, the simulation software that was used is freely available.

**Table 9.** Simulation engine software.

| Name | Type | Used in Papers |
|---|---|---|
| CloudSim | toolkit for simulating cloud computing infrastructures and services | [30,33,39,40,45,52,56,60,68,71,73,75,79,88,95,96,108,112,126,135] |
| CloudSimDisk | CloudSim module for simulating energy-aware storage in cloud systems | [96] |
| ICan Cloud | platform aimed to model and simulate cloud computing systems | [106] |
| iFogSim | resource management simulation toolkit for IoT, edge, and fog Computing Environments | [82,100] |
| ns2/ns3 | networking simulator tools | [51,97] |
| Mininet | network emulation and testing tool | [93] |
| OMNeT | C++ simulation library and framework, primarily for building network simulators | [106] |
| PureEdgeSim | simulation framework for performance evaluation of cloud, edge, and mist computing environments | [50] |
| SimPy | Python library for event-driven simulations | [83] |
| WorkflowSim | workflow simulator supporting large-scale scheduling, clustering and provisioning | [94] |

5.7.3. Programming Languages, Libraries, and Frameworks Used

Table 10 depicts numerical, scientific, and model creation frameworks and libraries used for implementing the proposed solution of each work. MATLAB was the software of choice for ten papers. Likewise, R is used in four papers. Python frameworks like Keras, Pytorch, Tensorflow, and scikit appear also, although in a smaller scale than expected, given the domination of Python as an ML language. A number of other more specific packages and libraries are mentioned, typically with more limited appearance in the overviewed works.

**Table 10.** Numerical and modeling frameworks and libraries.

| Name | Type | Freely Available | Used in Papers |
|---|---|---|---|
| ADTK | package [1] | Yes | [123] |
| CUDA Toolkit/ cuDNN | development environment for creating GPU-accelerated applications | Yes | [55,136] |
| E-GraphSAGE | network intrusion detection solution [2] | Yes | [115] |

**Table 10.** *Cont.*

| Name | Type | Freely Available | Used in Papers |
|---|---|---|---|
| Imbalanced-learn | Imbalanced classes classification extension for scikit-learn | Yes | [116] |
| jMetalPy | optimization library | Yes | [55] |
| Keras | framework | Yes | [53,70,121,122,129] |
| MAMLS | ML development environment [3] | No | [62] |
| MATLAB | scientific computing and development environment | No | [29,65,69,75,80,98, 109,111,114,119,130] |
| NumPy | scientific computing library | Yes | [53,129] |
| Pandas | data analysis library | Yes | [53,128,129] |
| PyGMO | optimization library | Yes | [55] |
| PyTorch | deep learning framework | Yes | [74,77,103,115,120, 127,130,136] |
| River | ML library for streaming data | Yes | [70] |
| Ryu | component-based software defined networking framework | Yes | [93] |
| scikit-learn | ML library | Yes | [53,128,129] |
| scikit-multiflow | ML library for streaming data | Yes | [70] |
| Spark | big data and ML platform | Yes | [44] |
| TensorFlow | deep learning framework | Yes | [70,84,116,121,122, 128,130] |

[1] Anomaly Detection Toolkit (ADTK) is a Python package for unsupervised, rule-based time-series anomaly detection. [2] E-GraphSAGE is a solution that uses graph neural networks to detect network intrusions in IoT networks based on flow records. There is a PyTorch-based implementation of E-GraphSAGE publicly available. [3] Azure Machine Learning Studio (MAMLS) is a GUI-based integrated development environment for constructing and operationalizing machine learning workflow on Azure.

In terms of general programming languages, although not listed in the table, the two most popular choices were Python and Java. Python was used in more than 20 measured cases, while Java was used in 6 papers, without counting the papers that use Java-based tools like CloudSim mentioned in the previous section. C, C++, and C# were also used in approximately six cases, usually combined with frameworks and libraries.

#### 5.7.4. Benchmarking and Load Generation Tools

In this section, we highlight a suite of tools (Table 11), including extensions of existing benchmarks, load generation tools, or elementary test applications that were used for benchmarking the performance and scalability of various cloud-based systems. In general, load generation is a critical step in any performance analysis and a common source of errors in the experimentation process. Hence, significant effort needs to be given on the way a workload is designed, applied, and validated on a given performance test. The existence and usage of helper tools for this purpose is therefore very significant. All benchmarking tools that were used are freely available.

**Table 11.** Benchmarking tools.

| Name | Type | Used in Papers |
|---|---|---|
| Bench4Q | extension of benchmarking tool TPC-W supporting QoS-oriented tuning of e-commerce servers | [131] |
| DeathStar Bench | benchmark suite for cloud microservices | [47,132] |
| JMeter | load testing Java tool | [59] |
| Locust | load testing Python tool | [55] |
| Online Boutique | microservice-based application with load testing capabilities | [47,132] |
| Yahoo! Cloud Serving Benchmark (YCSB) | program suite for evaluating retrieval and maintenance capabilities of computer programs | [135] |

*5.8. Datasets*

A large proportion of the papers used publicly available datasets, while others created synthetic datasets using appropriate tools. Tables 12–16 summarize publicly available and synthetic datasets and their use in relevant research papers. Publicly available datasets have been organized into four categories: application data (as a real-world workload for the services) in Table 12, network load and security data (Table 13), resource load (Table 14), and service workloads (including website and service request patterns) in Table 15. These are used typically to evaluate cases of simulations mentioned in Section 5.7.2 as well as inputs for the various validation strategies mentioned in Section 5.3.

Network load datasets are primarily used in network experiments focused on network traffic forecasting, and mainly contain traffic dumps, traffic logs, and network device telemetry data. Network security datasets are primarily used in network experiments focused on detecting network anomalies and recognizing network threats, and mainly contain labeled network activity, normal and malicious.

Resource load datasets are primarily intended to demonstrate realistic resource load, typically from traces of existing cloud providers and services (Alibaba, Google, Azure, etc.). They commonly contain aspects such as number of VMs, cores used, CPU/Memory usage, and more. While VM-based workloads exist, potentially, further datasets could be created that are more indicative of specific cloud-based services (e.g., cloud object storage services, messaging systems patterns, etc.). This is a current gap in the process that could help guide the creation of suitable extensions in these available data collections. An example of a workload trace for a very specific type of cloud service is the Azure Cloud Functions dataset (https://github.com/Azure/AzurePublicDataset/blob/master/AzureFunctionsDataset2019.md) (accessed on 14 February 2025) [137], that captures the individual characteristics of serverless workloads and has helped drive research in FaaS systems in the previous years. Indicatively, it has been cited more than 700 times in the years from its publication in 2019 up to 2025.

In the case of web traffic data, datasets are primarily used in experiments focused on cloud service auto-scaling and workload forecasting techniques, and contain web pages traffic traces, some from as far back as 1995. Further datasets could also be helpful that are tailored to usage patterns of applications more commonly met in cloud environments (e.g., AI training and inference, media streaming, IoT device feeds, etc.). Capturing the specific usage patterns of more modern and cloud-oriented applications could be critical for optimizing the management schemes of the latter.

Tables 12–15 are populated with dataset details, such as content, download link, in which paper each dataset was used, as well as inner level of data details that may aid researchers in selecting the appropriate data source for their experimentation. The existence of these as well as their categorization can significantly speed up both the selection process by a researcher, based on the scope of their research, as well as the experimentation itself. Furthermore, it enables direct comparison between different management approaches that are based on the same dataset.

**Table 12.** List of publicly accessible datasets used in papers for application-level automation.

| Dataset | Used in Papers |
|---|---|
| Chicago Taxi Trips (https://data.cityofchicago.org/Transportation/Taxi-Trips-2013-2023-/wrvz-psew/about_data) (accessed on 14 December 2024)<br>Used as application workload in examining the proposed microservice management framework. | [55] |
| COCO (https://cocodataset.org/) (accessed on 29 January 2025) [138]<br>Images used in self-adaptive application considering image characteristics during analysis as part of the proposed power saving mechanism. | [64] |
| COVID-XRay-5K (https://github.com/shervinmin/DeepCovid) (accessed on 2 October 2024)<br>Used as sample labeled storage data in proposed smart prefetching capabilities of a distributed file system. | [98] |
| EUA (https://github.com/swinedge/eua-dataset) (accessed on 2 November 2024) [139]<br>Cell base stations' location data used in proposed solution for application placement in an edge computing environment. | [50] |
| LandSat8 satellite images (https://earthexplorer.usgs.gov/s) (accessed on 26 October 2024) [140]<br>Sat images used as big data source in proposed adaptive data delivery method for solving data movement and processing bottlenecks in inter-site edge–fog–cloud systems. | [81] |
| LCTSC (https://www.cancerimagingarchive.net/collection/lctsc/) (accessed on 5 November 2024) [141]<br>Medical imagery dataset used in an edge–fog–cloud pipeline to measure and mitigate bottlenecks during offloading. | [81] |
| MDT-NJUST (https://github.com/YinLu-NJUST/MDT-2023) (accessed on 2 November 2024)<br>Contains real-world mobile devices trajectories in form of time-stamped GPS information. Used as a benchmark procedure source data to justify the performance of a mobility-aware off-and-downloading task algorithms in mobile edge computing. | [54] |
| New York City Taxi Trip Data (2010–2013) (https://doi.org/10.13012/J8PN93H8) (accessed on 22 September 2025) [142]<br>Taxi trip location data used to generate data stream workload for evaluation of a fog stream processing autoscaler. | [36] |
| The MIMIC-III clinical database (2017) (https://www.physionet.org/content/mimiciii/1.4/) (accessed on 11 December 2024)<br>Different IoT tasks (from sensor data collection up to cloud data ingestion) time-series data used as incoming load to evaluate the proposed adaptive IoT workflow management architecture. | [27] |
| UCI Heart Disease (https://archive.ics.uci.edu/dataset/45/heart+disease) (accessed on 27 January 2025)<br>Labeled medical data used to drive patient processing offloading based on patient criticality estimation. | [72] |

**Table 13.** List of publicly accessible datasets used in papers for network load and security automation. The majority of the datasets include Pcap capture files (https://www.ietf.org/archive/id/draft-gharris-opsawg-pcap-01.html) (accessed on 18 October 2024) with malicious activities. Pcap format is an industry standard used to capture and share information about any threat or network event. A Pcap file includes a series of packet records; each record represents a packet captured from the network, along with a timestamp and the length of the packet.

| Dataset | Used in Papers |
| --- | --- |
| Real mobile network traffic data (https://github.com/JinScientist/traffic-data-5min) (accessed on 18 October 2024)<br>Features data usage from a real mobile network cell in averages of 5 min intervals [143]. | [134] |
| ToN-IoT (https://research.unsw.edu.au/projects/toniot-datasets) (accessed on 14 December 2024)<br>Contains heterogeneous data sources collected from IoT and IIoT sensors and organized in four categories: raw datasets, processed labeled datasets, train test datasets (samples from the dataset with normal and malicious data), labeled hacking events, and statistics regarding the dataset [144]. | [115] |
| BoT-IoT (https://ieee-dataport.org/documents/bot-iot-dataset) (accessed on 26 January 2025) [145]<br>Raw network packets (Pcap files) created by tshark tool and incorporates a combination of labeled normal and abnormal traffic. | [115,125,126] |
| CIC-IDS (https://www.unb.ca/cic/datasets/ids-2018.html) (accessed on 21 November 2024) [146]<br>Features statistics in forward/backward direction including total/min/max/average/standard deviation of packet size. | [114,122,123,129,130] |
| IDE2012 (https://www.unb.ca/cic/datasets/ids.html) (accessed on 21 November 2024) [147]<br>Contains 7 days of network activity, normal and malicious. Dataset consists of labeled network traces, including full packet payloads in Pcap format. | [119] |
| N-BaIoT (https://www.kaggle.com/datasets/mkashifn/nbaiot-Dataset) (accessed on 25 January 2025)<br>Contains a rich set of 115 features extracted from real network traffic data, gathered from 9 commercial IoT devices authentically infected by Mirai and BASHLITE. | [130] |
| NSL-KDD (https://github.com/HoaNP/NSL-KDD-DataSet?tab=readme-ov-file) (accessed on 4 November 2024)<br>Contains labeled normal and attack traffic intrusion detection data. Features: duration, protocol, service, src/dst bytes, num of failed logins, su attemps, num failed logins, and many more. | [114,118,125,127] |
| UNSW-NB15 (https://research.unsw.edu.au/projects/unsw-nb15-dataset) (accessed on 21 November 2024)<br>Contains raw traffic Pcap and CSV files with nine types of attacks. | [118,119,130] |
| X-IIoTID (https://github.com/Alhawawreh/X-IIoTID) (accessed on 5 October 2024) [148]<br>Contains 68 features (including three security characterization label levels) extracted from network traffic, system logs, application logs, device's resources (CPU, input/Output, Memory, and others), and commercial intrusion detection system logs. | [122] |

**Table 14.** List of publicly accessible datasets used in papers for resource load automation.

| Dataset | Used in Papers |
| --- | --- |
| Alibaba cluster traces (https://github.com/alibaba/clusterdata) (accessed on 5 November 2024) [149]<br>Features: task id, job id, start time stamp, end stamp, machine id, container id, cpu avg max utilization, memory avg max utilization, cpu requested, memory requested, and more. | [48,74,78,88,107,135] |

**Table 14.** *Cont.*

| Dataset | Used in Papers |
| --- | --- |
| ASD (https://github.com/zhhlee/InterFusion/blob/main/data/DatasetDescription.pdf) (accessed on 18 October 2024) [150]<br>Contains 12 different server logs, each of which has 19 metrics characterizing the status of the server (including CPU-related metrics, memory-related metrics, network metrics, virtual machine metrics, etc.); | [120] |
| Azure Cloud 2017 trace (https://github.com/Azure/AzurePublicDataset) (accessed on 25 January 2025)<br>Features: timestamp VM created, VM deleted, count VMs created, VM id, cpu avg max utilization, VM category, VM memory, and more. | [74] |
| Azure Functions Dataset (https://github.com/Azure/AzurePublicDataset/blob/master/AzureFunctionsDataset2019.md) (accessed on 14 February 2025)<br>Features: function invocation counts and triggers, function execution time distributions, application memory allocation distributions, and more. | [74] |
| Bitbrains workload traces (http://gwa.ewi.tudelft.nl/datasets/gwa-t-12-bitbrains) (accessed on 5 November 2024) [151]<br>Contains 7 performance metrics per VM, sampled every 5 min: number of cores provisioned, the provisioned CPU capacity, CPU usage, the provisioned memory capacity, actual memory usage, disk I/O throughput, and network I/O throughput. | [46,67,128] |
| EMOS (https://github.com/FudanSELab/train-ticket) (accessed on 27 October 2024) [152]<br>Contains the status of 41 microservices, while faults were injected. Monitors four representative metrics, including CPU usage, RAM usage, Net out, and Net in. | [120] |
| GoCJ (https://data.mendeley.com/datasets/b7bp6xhrcd/1) [153] (accessed on 25 November 2024)<br>Contains jobs in terms of Million Instructions (MI) derived from the workload behaviors witnessed in Google cluster traces. | [30] |
| Google cluster workload traces (https://www.researchgate.net/profile/Auday-Al-Dulaimy/post/Are-there-any-datasets-for-cloudSim/attachment/59d61de379197b807797be3e/AS%3A273823268573184%401442295962733/download/Google+cluster+usage+traces.pdf) (accessed on 25 January 2025)<br>Cluster workload data, consisting of cluster jobs and tasks data. Features: timestamp, job id, user id, CPU/memory/disk space/disk I/O time resources, machine ID, and more. Includes details on machine capabilities (CPU, RAM etc.). | [39,43,46,78,89,102,103,113,135] |
| Kaggle Process Workload Dataset (https://surli.cc/cfpxsm) (accessed on 5 November 2024)<br>Contains jobs with features: burst time, arrival time, preemptive, and resources. | [104] |
| MBD (https://github.com/QAZASDEDC/TopoMAD) (accessed on 12 December 2024) [154]<br>Workload from a big data 5-node cluster. Contains randomly injected faults for CPU, network, and application levels and observations of 26 monitored metrics (CPU, disk, memory, network, and process) for the reaction of the cluster in these faults. | [120] |
| NASA iPSC (https://www.cs.huji.ac.il/labs/parallel/workload/l_nasa_ipsc/) (accessed on 18 October 2024)<br>Contains three months worth of sanitized accounting records for the 128-node iPSC/860 hypercube. Features: user, job, number of nodes, runtime, start date, start time, special entries about system status, duration and more. | [68,73,100] |
| NEP real-world edge workload (https://github.com/xumengwei/EdgeWorkloadsTraces) (accessed on 21 October 2024) [155]<br>Contains workloads traces of edge sites of China's largest public edge platform during June 2020. CPU, memory, storage, RTT, bandwidth traces at VM, and physical node granularity. | [77] |

**Table 14.** *Cont.*

| Dataset | Used in Papers |
|---|---|
| PlanetLab (https://github.com/beloglazov/planetlab-workload-traces) (accessed on 18 October 2024)<br>Ten-day real workload data included in the CloudSim framework. Contains traces of mean CPU utilization measured every 5 min of more than 1000 VMs running on thousands of servers in about 500 different locations globally. | [56,58,89] |
| Prediction dataset for cloud workload (https://github.com/vijayant123/Predicting-Cloud-Workload-Using-ANN/blob/master/Cloud_dataset.csv) (accessed on 5 November 2024)<br>Contains labeled cloud workload data such as Timestamp, CPU cores, capacity provisioned, usage, Memory usage, Disk read/write throughput, network usage, and more. | [101] |
| HPC2N workload (https://www.cs.huji.ac.il/labs/parallel/workload/l_hpc2n/) (accessed on 18 October 2024)<br>Three and a half years of HPC log records in Maui format (https://docs.adaptivecomputing.com/maui/trace.php) (accessed on 21 October 2024). Includes name of job, num of nodes and tasks requested, max allowed job duration, job completion state, timestamp for submitted job, job execution start, job completion, and many more. | [73] |

**Table 15.** List of publicly accessible datasets used in papers for Web Traffic Workloads automation.

| Dataset | Used in Papers |
|---|---|
| World Cup 98 Web Server (https://github.com/chengtx/WorldCup98) (accessed on 18 October 2024) [156]<br>Features: timestamp, clientID, objectID, size, method, status, type, server. | [41,53,95] |
| NASA Dataset (1995) (https://ita.ee.lbl.gov/html/contrib/NASA-HTTP.html) (accessed on 18 October 2024)<br>Features: host, timestamp, request, HTTP reply code, reply bytes. | [29,45,53] |
| Wikipedia article pageviews (https://wikimedia.org/api/rest_v1/) (accessed on 5 Novermber 2024)<br>Provides access to Wikipedia access data, including pageviews, unique devices, edited pages, editors, edits, registered users, bytes difference, media requests, and more. | [42] |
| Workload traces of Saskatchewan server (https://ita.ee.lbl.gov/html/contrib/Sask-HTTP.html) (accessed on 2 October 2024)<br>Features: host, timestamp, request, HTTP reply code, reply bytes. | [29,38] |

**Table 16.** List of synthetic datasets.

| Dataset Type | Used in Papers |
|---|---|
| Pre-constructed synthetic data | [35,45,48,52,70,76,82,94,104,105,108,116,121,132,133] |
| Synthetic data generated during the experiments | [26,28,34,37,47,51,57,59–61,65,66,69,71,75,79,80,83–87,90–93,96,97,106,109–111,124] |

## 6. Discussion, Conclusions, and Future Research Directions

This study focused on recent research work on both Cloud Design and Cloud Operation Patterns, analyzing 111 works from the perspectives mentioned in the defined taxonomy (pattern's prime goal, feedback loops, adaptation strategy, experiment validation methodology, runtime automation level, pattern's decision-making approach, software, tools, and datasets). The answers to the posed research questions in Section 3 have been described across the subsections of Section 5 and are summarized and extended below.

RQ1: Which are the patterns' objectives in cloud automation?

As indicated in Section 5.1, performance seems to have an overwhelming dominance when it comes to patterns (72%), followed by cost (26%). Performance includes placement optimization, network performance, response time reduction, or in general, QoS, as well as network improvement (bottleneck detection and SLO/QoS violation mitigation), while cost includes energy consumption, deployment, and operating costs. More than ten percent (16%) of the reviewed works deal with security issues of cloud computing, such as security anomalies and DDOS detection. Aspects such as availability and fault tolerance (9%) have not been sufficient investigated, areas that have a strong impact in complex and distributed cloud environments. Maintainability patterns could also aid in this direction, especially given that frequent changes needed in today's speedy development environments can result in errors.

RQ2: What are the scopes of control (i.e., target of regulation) of self-adaptive patterns?

Regarding control scope, the major percentage of research work investigates some form of resource management (infrastructure resizing, offloading, etc.) for an application, as indicated in Section 5.3, in contrast to the minor percentage that investigates dynamic adaptation of application architectures. There are a few promising approaches that offer even more fine-grained management by delving into the context of a specific application request. Although these approaches are less generic, they could offer strategic advantages on the context understanding of resource management for a given application.

RQ3: What are the approaches used for decision-making?

As depicted in Section 5.6, machine learning constitutes the primary decision-making approach in the researched publications. Additionally, time-series analysis and historical data processing are frequently employed, often in conjunction with machine learning techniques. Optimization and probabilistic algorithms are utilized in numerous instances, while some studies adopt simpler computational methods.

About half of feedback loop processes rely on reactive category (53%), while the rest uses a proactive approach (Section 5.2). Reactive solutions are generally lightweight and react to current state changes. In contrast, proactive approaches have the benefit of predicting future anomalies, thus preparing a priori for anticipated changes, but on the other hand, they may exhibit error in these predictions. This error may accumulate with the error in the process of determining the corrective action.

Regarding patterns' automation level (Section 5.5), in most of the cases, the "continuous analysis/training continues usage" category is preferred (84 cases), while the "one-off training/continuous usage" category used in 26 cases. The "one-off training/one off usage" category is utilized in two cases. These findings depict the preference for fully dynamic mechanisms, both in training and in usage.

RQ4: Which software and tools were used to create the management mechanisms in the reviewed work?

CloudSim was the most popular solution for developing and validating the proposed management mechanisms, followed by MATLAB and ML libraries and frameworks for Python such as Keras and TensorFlow, as depicted in Sections 5.7.2 and 5.7.3. There are several other libraries, frameworks, and tools that were used in individual cases. Regarding general language usage, Java and Python are the dominant cases.

RQ5: Which methods, datasets, and tools were used for experiment validation?

In the majority of the papers, experiments were based on simulations with tools such as CloudSim or MATLAB, while, to a lesser extent, real-world observations took place, as indicated in Section 5.4. Simulation software provides flexibility, cost reduction, and minimizes external factors that could influence results. However, it may also miss the parameters of real-world setups or the dynamic nature of distributed environments. Usage of real-world datasets in the simulation can help alleviate this aspect.

Running experiments on actual private or public cloud seems to be the minority of the overviewed work, as recorded in Section 5.7.1. Kubernetes orchestration system was the preferred software for real-world experiments, running either in private or public cloud like AWS, GCP, and Alibaba. In few cases of real-world experiments, specialized benchmarking and load testing software was used (Section 5.7.4).

A large proportion of the articles used publicly available datasets, while others created synthetic ones (Section 5.8). An effort was made to document these and group them based on their high-level scope (application, network, resource, traffic). A significant amount of open data exists; however, their usage for a given research goal is something that is difficult to evaluate. Thus, the documentation of their contents in this work can ease the effort for future researchers in order to identify the appropriate data source, as well as find similar works that have used them for comparison purposes.

The usage of different datasets (in the same category, e.g., for resources) seems to be somewhat fragmented. Many different ones are used, with no single dataset to be considered as a kind of industry or domain standard. Traces released from public cloud providers could play this role in some of the cases. On the same topic, a very limited number of papers are using benchmarks, which could be a way to standardize the experimentation process. Finally, from a workload trend perspective, datasets could be more specialized in relation to specific cloud service types or more modern, cloud-based applications.

*Additional Conclusions*

In addition to the research question analysis, the following general conclusions were extracted from this study. There is a significant rise in the works of recent years, compared to the period before 2023. For years prior to those included in this review (2018–2025), a brief search revealed 34 relevant research papers published between 2010 and 2017. Thus, the topic of cloud and application management automation is expected to continue being of interest, especially with the advent of more complex approaches like agentic AI.

Although close to a hundred papers were studied, only seven of them appear to provide accessible source code. This limited availability hinders the reproducibility and further development or comparison of research findings. Furthermore, the number of papers mentioned in the tables in Section 5.7 (on used tools) is significantly lower than the total number of papers reviewed. This indicates a low number of papers that explicitly mention the tools and libraries they use, which is another factor limiting reproducibility.

More than 70 percent of the researched work is in journals, which can indicate the complexity of the given domain, as journal articles typically delve into specialized topics and more in-depth analysis and experimentation. This level of complexity indicates the significant expertise and system-wide knowledge that is required and may span from system setup (typically from a systems engineer) to decision mechanism creation (typically from a data scientist) and experiment/validation organization (typically from a performance engineer). Given that this type of expertise combination is rare, it is imperative to work as a community towards more standardized experimentation means, including potential deployment and configuration templates.

The small amount of solutions at the application level points to a future research area in the field of cloud self-adaptive applications. Researchers and cloud-native application developers, exploiting the new capabilities offered by artificial intelligence [157], can pursue the creation of intelligent adaptation mechanisms within applications. These mechanisms can enable self-adaptive characteristics directly within the application structure. This adaptation may involve switching between parametric application architectures, as well as embedding the logic needed to decide on the transition between the available configurations. Such applications, which are more "liquid" and flexible, could lead to inherent and multimodal self-adaptive capabilities.

## 7. Proposed Future Directions

Concluding the overview findings, we propose a collaborative application and system-level design (Figure 3) that constitutes a potential blueprint for cloud self-adaptive applications. The building blocks of this architecture are also annotated with respect to their relation to the research questions (RQs) analyzed in the previous sections. The application and the system layer maintain a vital two-way data exchange with the aim of reacting to both the application and the system changes (related to RQ3). This exchange is not mandatory, but it can help the application have a wider context of its execution within the system, as well as the system to understand better the application context when it needs to conclude on its own decisions (placement, scheduling, etc.).

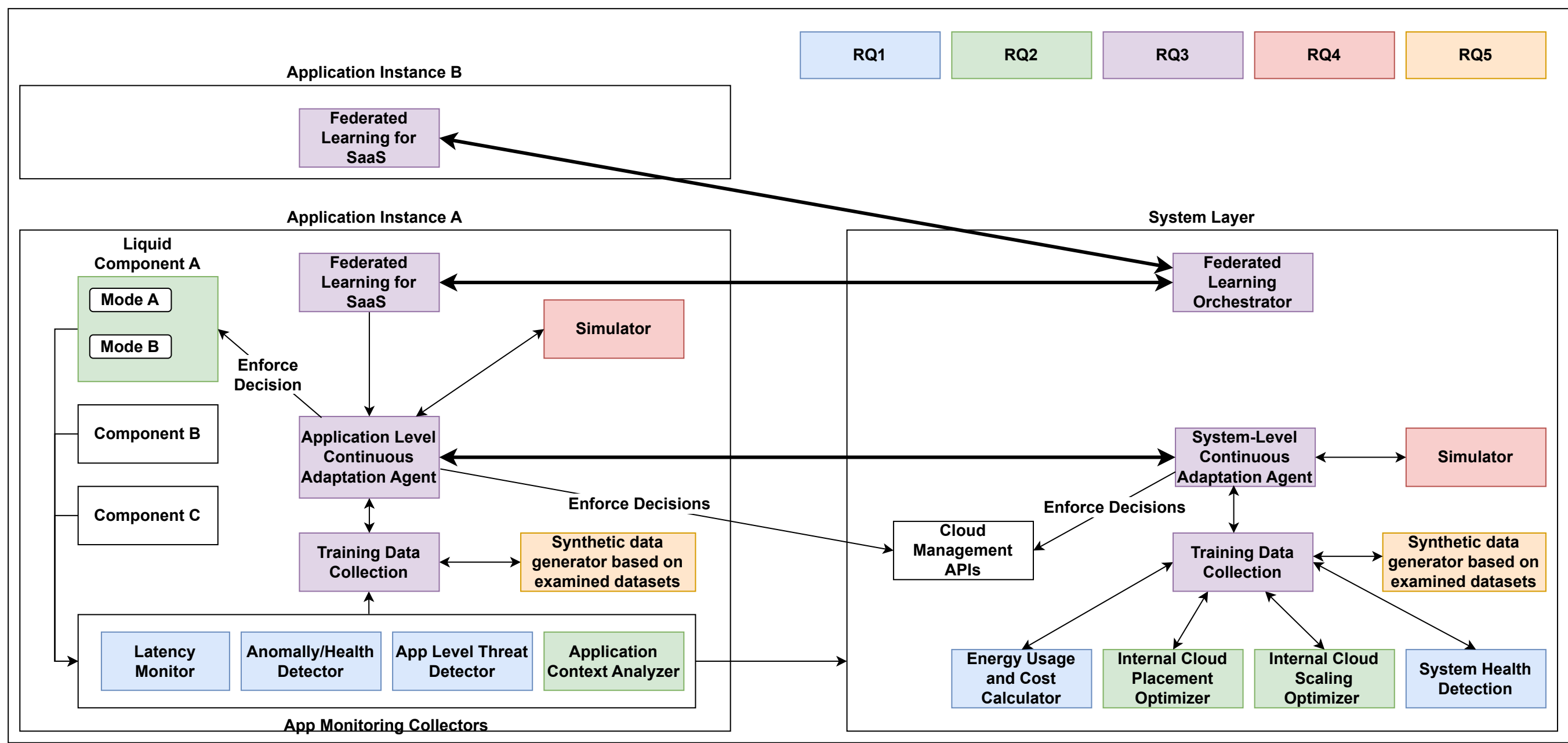

**Figure 3.** Generic app-system-level design based on the outcomes of the RQs.

For the application layer, the main proposal refers to applications that are enriched with intelligent and embedded adaptation mechanisms, adding to them self-adaptive characteristics directly within the application structure in synergy with the system layer orchestrator. This solution includes the ability to switch between diverse component (or "liquid") modes and embeds the logic needed to decide on the transition between the available implementations.

The central entity is the Application-Level Adaptation Agent (RQ3), which is fed with data by four monitoring components, the Latency Monitor (RQ1), the Anomaly/Health Detector (RQ1), the Application-Level Threat Detector (RQ1), and the Application Context Analyzer (RQ2) components, as well as available data from publicly available datasets (RQ5). Based on the training of an according model, the Agent can then decide on a corrective action. Simulation of the action can be performed via the Simulator (RQ4) to evaluate the efficiency of a new configuration solution before applying it to production. The decision, related to either resources assigned to the application, of other configuration parameters or which form the application should obtain (from the "liquid" ones) is enforced based on available application or system APIs.

An added value component in this case refers especially to Software as a Service (SaaS) applications. One of the forms of SaaS refers to the deployment of discrete but identical (in structure) application instances, one for each tenant/account. This means that an application management model for one instance could also be applied to other instances. It also means that monitoring data collected from many instances, with potentially different usage patterns would help in the more effective training of the Application Agent. However, given that the instances relate to different tenants, sensitive or business-related information could exist within the training logs; thus, suitable security and privacy mechanisms should be put in place before collaborating on these training data.

This is the case of federated learning [158], through which the SaaS instances can collaborate without actually releasing their data. Different modes of collaboration may be applied. In a more loosely coupled collaboration, each instance can locally train their Agent and there is collective (or ensemble) inference on a desired action that is applied in a voting or averaging fashion. For example, if the Agent of Instance A decides that it needs two replicas of a component to deal with an incoming traffic surge, it can inquire Instance B's model and average out the number of needed replicas. A more collaborative method refers to collective training, in which each instance starts the Agent training and after a few epochs, weights and parameters are shared between the participants in the federation. This process iterates until all epochs are complete. Thus, the instances create a collaborative model based on every tenant training data, without actually sharing them. To avoid information leakage from weights and model parameters, federated learning can also be combined with homomorphic encryption approaches [159]. Federated learning approaches typically need the existence of a more centralized orchestrator, residing at the system level or at the location of the entity offering the SaaS solution (SaaS provider).

For the system level, the according System-Level Adaptation Agent (RQ3) prepares the training data collection with the help of the Data Generator (RQ5) and feeds them to four discrete optimization components: the Energy Usage and Cost Calculator (RQ1), the Internal Cloud Placement Optimizer (RQ2), the Internal Cloud Scaling Optimizer (RQ2), and the System Health Detection (RQ1) components. In a similar fashion to the Application Layer, the Agent can decide on a corrective action based on a according model; decisions can also be enforced by Cloud Management APIs. The system-level models can be enriched with parts of the application context to improve their efficiency. Moreover, the Simulator (RQ4), core part of the system mechanism tests subsequent decisions of the optimizers.

The proposal for self-adaptive cloud application and system design emphasizes the creation of systems capable of autonomously adjusting their behavior in response to dynamic environmental conditions, fluctuating workloads, evolving user needs, and even security threats. This paradigm shift moves beyond traditional static designs towards architectures that incorporate multiple implementation modes as well as dynamic configuration for the selection of the appropriate mode or parameter.

However, significant challenges remain for future elaborations of this architecture. These include handling transitions between application modes, managing the heterogeneity of components, and coordinating adaptations across the application.

**Author Contributions:** Conceptualization, A.A. and G.K.; methodology, A.A. and G.K.; investigation, A.A.; data curation, A.A.; writing—original draft preparation, A.A.; writing—review and editing, G.K.; supervision, G.K. All authors have read and agreed to the published version of the manuscript.

**Funding:** Part of the research leading to the results presented in this paper has received funding from the European Union's funded Project HUMAINE under grant agreement No. 101120218.

**Data Availability Statement:** Data analyzed in this study were derived from resources available in the public domain, as described in detail in Section 5.8.

**Conflicts of Interest:** The funders had no role in the design of the study; in the collection, analyses, or interpretation of data; in the writing of the manuscript; or in the decision to publish the results.

## Abbreviations

The following abbreviations are used in this manuscript:

| | |
|---|---|
| ARIMA | Autoregressive Integrated Moving Average |
| DDoS | Distributed Denial of Service |
| LSTM | Long Short-Term Memory (LSTM) |
| MAPE-K | Monitor–Analyze–Plan–Execute over a shared Knowledge |
| QoS | Quality of Service |
| RQ | Research Question |
| SLO | Service-Level Objectives |